\documentclass[twocolumn,prd,showpacs,preprintnumbers,amsmath,amssymb]{revtex4}

\newcommand{\dt}{\ensuremath{\, \mathrm{d}}}
\newcommand{\dti}{\ensuremath{\, \mathrm{d}\kern-0.1em{}}}

\newcommand{\set}[1]{\left\{#1\right\}}

\newcommand{\defin}{\mathrel{\mathop:}=}
\newcommand{\eqrf}[1]{Eq.~(\ref{#1})}
\newcommand{\eqsrf}[1]{Eqs.~(\ref{#1})}

\newcommand{\hide}[1]{\relax}

\newcommand{\punctsp}{ }

\usepackage{tensor}
\usepackage{hyperref}

\begin{document}

\title{The Reciprocal of the Fundamental Theorem of Riemannian Geometry}

\author{H\'{e}ctor H. Calder\'{o}n}
\affiliation{Department of Physics and Astronomy, Carleton College, Northfield, Minnesota, 55057}

\email{calderon@carleton.edu}
\date{\today}
\begin{abstract}
The fundamental theorem of Riemannian geometry is inverted for analytic Christoffel symbols. The inversion formula, henceforth dubbed Ricardo's formula, is obtained without ancillary assumptions. Even though Ricardo's formula can mathematically give the full answer, it is argued that the solution should be taken only up to a constant conformal factor. A procedure to obtain the Christoffel symbols out of unparameterized geodesics is sketched. Thus, a complete framework to obtain the metric out of measurements is presented. The framework is suitable for analysis of experiments testing the geometrical nature of gravity.
\end{abstract}
\maketitle

\section{Introduction}

The archetypical experiment in General Relativity consists in following the trajectory of tests particles and determining the presence of the gravitational field from the mutual deviation of these world lines. Postulating that test-particle trajectories are geodesics is not a trivial step. After all, a world line is deemed to be a geodesic only if its length is a extremum. The notion of length requires a metric. Therefore, the analysis of any experiment designed to test the strong equivalence principle should avoid making assumptions that are tantamount to having a metric available, even an unknown one. On the other hand, if the very existence of the metric is not in question, then there are methods to figure either the connection or the curvature out of the geodesics \cite{Synge2009, Pirani1956, Szekeres1965, Ciufolini1986}. From there, one can apply any of the formal methods available in the literature to find the metric \cite{Schmidt1973, Ihrig1975a}.

Physicists favor computing the curvature because the connection, being a collection of tensors but not a tensor itself, is usually not deemed to encode a physical object. Besides, if one had the connection, the curvature is just a few calculations away. Thus, finding the metric out of the connection or out of the curvature are equivalent steps in the search of the metric out of observations. Mathematically speaking however, the connection is at the center of the Fundamental Theorem of Riemannian Geometry: For a given metric, there is one and only one torsion-free connection that parallel transports the metric. In a coordinate system, the components of such connection is given by Christoffel's formula:
\begin{equation} \label{ChrDef}
\Gamma \indices {^\alpha_{\beta \gamma} } = \frac {1} {2} g^{\alpha \mu} \left( g_{\mu \beta , \gamma} + g_{\mu \gamma , \beta} - g_{\beta \gamma , \mu} \right)\punctsp.
\end{equation}

The Inverse of the Fundamental Theorem of Riemannian Geometry is the solution of the above partial differential equation in the metric when the Christoffel symbols and appropriate boundary conditions are given. Hereafter, we will shorten this as ``finding the metric''. This may include the calculation of the curvature as an intermediate step. With ``finding the Christoffel symbols'', we will refer to the step of computing the components of the connection out of trajectories in the coordinate system in which the trajectories are specified.

Breaking the calculations in these two steps is implicit throughout the literature. Perhaps the only counterexample is Singatullin's calculations of the metric out of properties of geodesics valid for static metrics \cite{Singatullin1978}.

Finding the metric is tackled informally by postulating the most general metric compatible with some simplifying symmetries (\emph{e.g.} spherical symmetry), computing the connection from such metric, and best-fitting it against the measured values. The main drawback of this procedure is that the existence of the metric is assumed \emph{a priori}. These analyses should not be used to test the strong equivalence principle for example.

Known to the author, there are two classes of procedures to find the metric. The algorithm devised by Schmidt \cite{Schmidt1973} requires computing the holonomy group \footnote{The holonomy group at a point $x$ is the group of linear transformations in the tangent space of $x$ determined by parallel transport along closed loops from $x$.} and all of its quadratic invariant forms. However as pointed out by Ihrig, this is rather only a formal solution on the account that it still involves solving the parallel-transport equations \cite{Ihrig1975}. Moreover, the emphasis on the holonomy group makes the problem unnecessarily difficult because the solutions are sought independently for each holonomy class (\emph{e.g.}, as done by Hall \cite{Hall1988} or Ghanam \cite{Ghanam2002}). Also, classifying the connections in clear-cut classes is incompatible with the fuzzy values associated with uncertainties.

The second algorithm to find the metric was originally presented by Ihrig \cite{Ihrig1975a}. It determines the metric using the curvature instead of the connection \cite{McIntosh1981, Hall1983, Edgar1991, Quevedo1992} and the solution is found with some restrictions (for example, when certain matrix is invertible, \emph{c.f.} Eq.~(14) in \cite{Edgar1991}). It is an attractive method in that there are no differential equations to solve. Still, this approach is not desirable to compute the metric uncertainty, because the restrictions are not guaranteed to be satisfied for all values inside of the interval of confidence (of the curvature). None of these two programs deal satisfactorily with the issue of boundary conditions.

This paper presents a direct method to find the metric without assuming any simplifying symmetries. It is only required that the Christoffel symbols are analytic in a geodesically connected patch \footnote{In a geodesically connected manifold, any two points can be connected by a geodesic. Path connectedness and completeness are necessary for geodesical connectedness.} and that the set of isometries are known at a point of such patch. The analyticity requirement is stronger than the usual smoothness condition \cite{Hall1988}, but it should not pose practical restrictions for ``real-life'' experiments. The solution works for spaces of any dimension and signature.

For the sake of completeness, we will review the problem of finding the Christoffel symbols in the next section. This will allow us to cover the whole procedure from measurements to metric with the same philosophy of avoiding extra assumptions. The main result is derived in the third section and analyzed in the following one. An example is provided in the appendix.

\section{The Christoffel Symbols}
The problem of finding the Christoffel symbols (or the curvature) out of trajectories of test objects has been studied before. Synge developed the five-point method \cite{Synge1960,Synge2009}, Pirani followed the trajectory of a cloud \cite{Pirani1956}, Szekeres proposed the gravitational compass \cite{Szekeres1965}, Ciufolini used Fermi-Walker transport \cite{Ciufolini1986}. In all of these works, the trajectory is known as a function of the affine parameter. Otherwise in those derivations, it would have been impossible to assert that some 4-vector is the (normalized) 4-velocity of the test particle. Moreover, as Quevedo pointed out in relation to the above cited methods \cite{Quevedo1992}, ``the measurement of the curvature is associated with an orthonormal tetrad comoving with an observer''. However, this implies an \emph{a priori} notion of orthonomality. In other words, one cannot assert that a tetrad is orthonormal until the metric has already been found. A similar case is made against using Fermi-Walker transport whose defining equation implicitly involves the metric.

The affine parameter might be available under some special cases and further information can be extracted \cite{Hall2007}. However, it is important to obtain the connection without assuming that the connection is a metric connection because alternative theories of gravity might differentiate \cite{Borunda2008} between affine geodesics and metric geodesics \footnote{Connections that cannot be derived from a metric produce affine geodesics. This is in contrast to metric geodesics whose connection is a metric one.}. In fact, dealing with geodesics to find the correct metric might not be trivial and errors might go unnoticed for generations \cite{Finn2008}.

If the ``metricity'' of the connection is not challenged, then the uniqueness of the Christoffel symbols has been proven for vacuum \cite{Hall2007}, for Friedmann-–Lema\^{i}tre-–Robertson-–Walker \cite{Hall2008a}, and for certain Einstein \cite{Kiosak2009} spacetimes. As shown by Marmo \emph{et al}, the requirement of being a metric connection is a very strong one: There are connections that can be derived from a Lagrangian but fail to be metric connections \cite{Marmo1990}. An important case of nonmetric connection is Newtonian mechanics \cite{McIntosh1981}. Thus, determining the ``metricity'' of a connection becomes a test of General Relativity.

This paper is concerned with the general experimental situation where each trajectory is characterized only by the coordinates of its events. That is, the experimentalist provides us with the unparameterized trajectories of test particles. Since the affine parameter is unknown, the Christoffel symbols cannot be read off directly from the geodesic equation
\begin{equation} \label{geodeqparam}
\ddot x^\alpha + \Gamma \indices {^\alpha_{\mu \nu} } \, \dot x^\mu \dot x^\nu = 0 \punctsp.
\end{equation}
Instead, one can parameterize the trajectories in terms of a coordinate, say $x^b$. \eqrf{geodeqparam} is then transformed using
\begin{align}
\dot x^a &= x \indices {^a_{\!,b\,} } \dot x^b \punctsp, \text{ and }\\
\ddot x^a &= x \indices {^a_{\!,bb\,} } \dot x^b \dot x^b + x \indices {^a_{\!,b\,} } \ddot x^b \punctsp,
\end{align}
where $x^a = x^a( x^b )$ is the $a$-th coordinate considered as a function of the $b$-th coordinate, $x \indices {^a_{,b}}$ indicates the derivative of $x^a$ respect to $x^b$, and $x \indices {^a_{,bb} }$ stands for the second derivative. There is no summation on repeated Latin indices. We have then
\begin{equation}
x \indices {^b_{\!,b} } = 1 \quad \text {and} \quad x \indices {^b_{\!,bb} } = 0
\end{equation}
and the geodesic equations for $a$ and $b$ become
\begin{align}
\ddot x^b + \Gamma \indices {^b_{\mu \nu} } x \indices {^\mu_{\!,b \,} } x \indices {^\nu_{\!,b \,} } \, \dot x^b \dot x^b = 0 \punctsp,\\
x \indices {^a_{\!,bb} } \, \dot x^b \dot x^b + x \indices {^a_{,b} } \ddot x^b + \Gamma \indices {^a_{\mu \nu} } x \indices {^\mu_{\!,b \,} } x \indices {^\nu_{\!,b \,} } \, \dot x^b \dot x^b = 0 \punctsp.
\end{align}
Assuming that the geodesic does not correspond to a particle standing still and that $x^b$ is not a constant of the motion, we can remove $\ddot x^b$ from the equations above and arrive to
\begin{equation} \label{findGamma}
x \indices {^a_{\!,bb} } + x \indices {^\mu_{\!,b\,} } x \indices {^\nu_{\!,b} } \left( \Gamma \indices {^a_{\mu \nu} } - x \indices {^a_{\!,b\,} } \Gamma \indices {^b_{\mu \nu} } \right) = 0\punctsp,
\end{equation}
which is a differential equation with $x^b$ as the independent variable and no reference to the affine parameter whatsoever.

Because of the parameterization $x^a = x^a( x^b )$ on the trajectories, \eqrf{findGamma} can be interpreted as a linear equation on the unknown Christoffel symbols. In four dimensions, each trajectory can potentially give 12 scalar equations by taking different values of $a$ and $b$ in \eqrf{findGamma}. Thus, one needs to consider at least four trajectories per point in order to find the 40 values of $\Gamma \indices {^\alpha_{\mu \nu} }$. It can be proven however that \eqrf{findGamma} does not determine the Christoffel symbols uniquely even if we used more than four trajectories trough the same point.

The non-uniqueness of the mapping from trajectories to connections does not come out as a surprise. After all, we have Levi-Civita's theorem \cite{Levi-Civita1986}: two connections $\Gamma \indices{^\alpha_{\beta \gamma } }$ and $\overline{\Gamma} \indices{^\alpha_{\beta \gamma } }$ have the same unparameterized geodesics if and only if their difference can be written as $\delta \indices {^\alpha_{\! \beta \,} } \xi_\gamma + \delta \indices {^\alpha_{\! \gamma \,} } \xi_\beta$, where $\xi_\gamma$ is an arbitrary covariant vector.

\section{The metric}
From \eqrf{ChrDef}, it is trivial to obtain a linear first-order partial differential equation on the metric components:
\begin{equation} \label{originaleqg}
  g_{\alpha \beta ,\gamma } = \Gamma\indices{^{\mu \nu }_{\alpha \beta \gamma }}\,g_{\mu \nu } \punctsp,
\end{equation}
where
\begin{equation*}
  \Gamma\indices{^{\mu \nu }_{\alpha \beta \gamma }} \defin \Gamma \indices{^\mu _{\alpha \gamma }}\delta \indices{^\nu _\beta} + \delta \indices{^\mu _\alpha} \Gamma\indices{^ \nu _{\beta \gamma }}\punctsp.
\end{equation*}
This equation can be easily solved by converting it into an ordinary differential equation. We can accomplish this by inverting the reasoning that defines a partial derivative as a special case of the directional derivative. We will solve \eqrf{originaleqg} along a path
\begin{equation} \label{patheq}
  x^\mu \defin {x_0}^\mu  + \lambda ({x_f}^\mu - {x_0}^\mu)
\end{equation}
from the point $x_0$ (where the isometries are known) to the point $x_f$ (where the metric is to be evaluated). Since the Christoffel symbols can now be interpreted as functions of the parameter $\lambda $, then \eqrf{originaleqg}, after multiplying both sides by ${\frac {{\dt x^{\!\gamma}}\!\!} {\dt \lambda}}$, becomes an ordinary differential equation:
\begin{equation} \label{ode}
   \frac{\dt g_{\alpha \beta }} {\dt \lambda } = \Gamma \indices{^ {\mu\nu }  _{\alpha \beta \gamma }} ({x_f}^\gamma  - {x_0}^\gamma ) g_{\mu \nu }\punctsp.
\end{equation}
It can be solved, say using Frobenius method, and evaluated at $\lambda  = 1$.

The expansions of $g\defin g_{\alpha \beta }$ and $\Gamma \defin \Gamma \indices{^\mu^\nu _{\alpha \beta \gamma }}( {x_f}^\gamma   -  {x_0}^\gamma  )$ around $x_0$ can be written as
\begin{align}
g &= \sum_{i = 0}^\infty a_i \lambda^i \punctsp, \text { and } \label{gseries} \\
\Gamma &= \sum_{i = 0}^\infty b_i \lambda^i \punctsp. \label{gammaseries}
\end{align}
The tensor indices have been removed to avoid clutter in the expressions. After replacing \eqsrf{gseries} and (\ref{gammaseries}) into \eqrf{ode} and gathering the powers of $\lambda $, we find
\begin{subequations} \label{RicardoFormula}
\begin{align}
a_0 &= g_0 \punctsp, \label{RFBC} \\%
a_i &= \frac {1} {i} \sum_{j = 0}^{i - 1} a_{i - 1 - j} \star b_j \punctsp, \label{RFPS} \\
g &= \sum_{i = 0}^\infty a_i \label{RFFS} \punctsp,
\end{align}
\end{subequations}
where $g_0$ is the metric at $x_0$ and the operation $\star$ is defined by $p \star q \defin p\indices{_{\mu \nu }}q\indices{^{\mu \nu }_{\alpha \beta }}$. \eqsrf{RicardoFormula} shall be named Ricardo's formula.

\section{Analysis}
Finding the metric has been studied extensively \cite{Cocos2005, Edgar1991, Edgar1992, Hall1983, Hall1988, Hall2006, Hall2007, Hall2008a, Ihrig1975a, Ihrig1975, Kiosak2009, Marmo1990, McIntosh1981, Rendall1988, Schmidt1973, Thompson1993, Vermeil1918, Ghanam2002, Quevedo1992}. These works provide a mixture of somewhat general algorithms and studies of existence and uniqueness. Ricardo's formula has the advantage of being quite universal. The only requirements are analytic Christoffel symbols in a geodesically connected patch. The analyticity prerequisite excludes pathological examples (\emph{e.g.} Eqs.(6)-(8) of \cite{Schmidt1973}). The demand of geodesical connectedness will take care of ambiguities arising from singularities.

While the convergence of \eqsrf{RicardoFormula} can be proven for any set of $n^2( n + 1 ) / 2$ analytic functions, where $n$ is the dimensionality of the manifold, it has already been proven that not any such set of functions corresponds to the Christoffel symbols from a metric (an example can be found in Eq.~(1) of \cite{Cocos2005}). Edgar \cite{Edgar1992} gave the necessary and sufficient conditions for a connection to be locally metric \footnote{A connection is locally metric at a point if there exists a neighborhood of such point where the connection can be derived from a metric.}:
\begin{equation} \label{integrabcond}
h \indices {_{\mu ( \alpha} } R \indices {^\mu_{\beta ) \gamma \delta} } = 0 \punctsp,
\end{equation}
where $h$ is the so-called metric candidate. In our case, $h$ is just $g$ of \eqrf{RFFS}. Atkins extended Edgar's results when he proved that an analytic locally metric connection on an analytic simply-connected manifold is metric \cite{Atkins2008}. Thus, the integrability condition of \eqrf{originaleqg} is \eqrf{integrabcond}. In practice, it is easier to just check that the connection is recovered using \eqrf{ChrDef}.

Now, let us focus on the uniqueness of the solution. Being a linear first-order partial differential equation with analytic coefficients, the Cauchy-Kowalevski theorem implies that \eqrf{originaleqg} should have a unique solution determined by the appropriate boundary conditions (see page 348 in \cite{Hormander1983}). However, the solution of \eqrf{ode} might be path dependent. We can prove that this is not the case by integrating \eqrf{ode} along a closed loop. The resulting right-hand side can be cast as the surface integral of $g\indices{_{\mu (\alpha}}R\indices{^\mu_{\beta) \gamma \delta }}$, which is known to vanish (the connectedness requirement avoids enclosing any singularity within the integration path). Therefore, $\oint\frac{\dt g} {\dt \lambda } \dt \lambda = 0$ and the metric defined by \eqsrf{RicardoFormula} is independent of the path chosen in \eqrf{patheq}.

For physicists, it is reassuring to know that the signature of the metric does not change from event to event in spacetime. In order to prove this, it suffices to compare the signature of the metric at $x_f$ and the signature of $g_0$. Since the metric given by \eqrf{gseries} is continuous, the signature at $x_0$ can be different from the signature at $x_f$ only if there exists at least one point in a path joining $x_0$ and $x_f$ where the determinant of the metric vanishes. But at such point, because of \eqrf{ChrDef}, the Christoffel symbols would cease to be analytic. Note that the constancy of the signature is a consequence of the analyticity requirement and it is valid for geodesically connected manifolds. In particular, it has nothing to do with the equations of motion that yielded the geodesics.

The procedure in the previous section has broader application to solve a larger class of partial differential equations. Also, there are implications in the handling of the boundary conditions. Both of these issues are out of the scope of this paper. For the purposes of this article, we note that the boundary conditions have been replaced by the knowledge of the set of isometries at $x_0$. This means that we know $g_0$, the actual value of the metric at $x_0$, used in \eqrf{RFBC} up to a positive factor. This factor propagates to all terms of the series of \eqrf{RFFS} because of \eqrf{RFPS}. Therefore, our solution is given up to a positive constant conformal factor. If, instead of using the set of isometries as boundary conditions, we used $g_0$ then the conformal factor would have been fixed.

In the literature \cite{Rendall1988, Hall2007}, the indeterminacy of the overall factor in the metric is usually taken to correspond to the freedom of choosing the unit of length. This is only partially correct. The coordinates of the events in the geodesics were determined with some measurement apparatus. The calibration of the apparatus settles the units.

What is missing is a way to relate the units for the various axis. For example, why do we stipulate that a light year along the $x$ direction is the same as a the light year along the $y$ direction or the same as a year along the $t$ direction? Such relationships need to be specified only at one point. The connection then translates the relationships to the rest of the patch. The relationships are statements about the measurement apparatus and they are independent of the Physics that determined the geodesics. For instance, Local Lorentz Invariance cannot be used to lift the indeterminacy.

The relationships gather in the set of isometries but not on the abstract group of isometries. In order to understand why we focus on the set of isometries, consider the role of $c$ on
\begin{equation}
\dt s^2 = -c^2 \dt t^2 + \dt x^2\punctsp.
\end{equation}
Different values of $c$ correspond to the speed of light being measured using different units. The set of isometries is not the same for each value of $c$ (in the same sense that the interval $[0,1]$ is different from the interval $[0,2]$), although the abstract group is the same. Pointwise, the value of the metric determines the set of isometries but the set of isometries decides the metric only up to a factor. In short, the measurement apparatus fixes the units and the set of isometries at the point from where the apparatus is calibrated.

The indeterminacy would be lifted if either the Physics or the manifold itself provided a fundamental length. General Relativity does not have one built in; and spacetime, being fundamentally featureless, cannot furnish one either.

The method in this paper has both advantages and disadvantages respect to the ones available in the literature. The most prominent difference is the use of coordinate-dependent quantities (\emph{e.g.} \eqrf{gammaseries}) to find the solution of a coordinate-independent equation ($\nabla g = 0$). Two rewards are worth of notice. First, the solution is general enough to cover the demands of Physics. Second, one can truncate the series to compute all of the metrics compatible with the geodesics up to the accuracy of the instruments.

The results of this paper open the possibility of further studies on the stability of the inversion procedure.  Such studies have been done for 3D Riemannian metrics \cite{Wang1999a} but there is no definitive answer for 4D pseudo-Riemannian ones.

\section*{Acknowledgments}
The author is deeply indebted to his advisor, William A. Hiscock, for his support and guidance. The bulk of the research for this paper was done during the author's stay at Montana State University -- Bozeman.

\appendix*
\section{}
Let us accept as hypothesis that an experimentalist has collected enough data to claim that the trajectories in a region of spacetime are given by
\begin{equation}
x^\mu = ( t,z ) = \left( t , v_z \, t + \frac {1} {2} \alpha \, t^2 + \frac {1} {6} \beta \, t^3 + \mathcal {O} \left( t^4 \right) \right) \punctsp, \label{ExParam}
\end{equation}
where $v_z$ is constant in a trajectory, $\alpha$ and $\beta$ are constants for all trajectories, and the fitting is good to the order shown by $\mathcal {O}$. For this parametrization, we get
\begin{align}
x \indices {^\mu_{\!t} } &= \left( 1, v_z + \alpha\, t + \frac {1} {2} \beta \,t^2 + \mathcal {O} \left( t^3 \right) \right)\punctsp, \\
x \indices {^\mu_{\!t t} } &= \left( 0, \alpha + \beta \,t + \mathcal {O} \left( t^2 \right)\right) \punctsp,
\end{align}
and \eqrf{findGamma} yields
\begin{multline}
\alpha +\beta \,t + x \indices {^\mu_{\!t \,} } x \indices {^\nu_{\!t \,} } \left( \Gamma \indices {^z_{\mu \nu}} - \right.\\ \left.\left( v_z + \alpha\, t + \frac {1} {2} \beta \,t^2 \right) \Gamma \indices {^t_{\mu \nu} } \right) = \mathcal {O} \left( t^2 \right)\punctsp.
\end{multline}
Since $v_z$ is a parameter that names the geodesic, the Christoffel symbols should not depend on it. Thus, the coefficients of all powers of $v_z$ must vanish in the last equation. This is equivalent to generate several equations for different values of $v_z$. After solving the resulting equations, one gets
\begin{subequations} \label{ExampleConnection}
\begin{align}
\frac {1} {2} \Gamma \indices {^t_{t t} } = \Gamma \indices {^z_{t z} } = \Gamma \indices {^z_{z t} } = \xi_1 + \mathcal {O} \left( t^2 \right)\punctsp,\\
\frac {1} {2} \Gamma \indices {^z_{z z} } = \Gamma \indices {^t_{t z} } = \Gamma \indices {^t_{z t} } = \xi_2 + \mathcal {O} \left( t^2 \right)\punctsp,\\
\Gamma \indices {^z_{t t} } = - \alpha - \beta \,t + \mathcal {O} \left( t^2 \right)\punctsp.
\end{align}
\end{subequations}
The components not shown vanish. The unknown functions $\xi_i$ make the covariant vector anticipated by Levi-Civita's theorem.

The parametrization in \eqrf{ExParam} implies that we should keep track of $\xi_i$ only up to $\mathcal {O} \left( (t,z)^2 \right)$.
We can then write
\begin{align}
\xi_1 &= \xi_{10} + \xi_{11}t + \xi_{12}z + \mathcal {O} \left( (t,z)^2 \right) \punctsp,\\
\xi_2 &= \xi_{20} + \xi_{21}t + \xi_{22}z + \mathcal {O} \left( (t,z)^2 \right)\punctsp,
\end{align}
where $\xi_{ij}$ are constants.

Let us accept as hypothesis that the rods and clocks used were such that if an element $u$ of the tangent space at $( 0, 0 )$ is acted by an element of the set $S$ defined below, then the length of $u$ remains constant.
\begin{equation}
S = \set {
\left(
\begin{array} {cc}
\cosh ( \psi ) & \frac {3} {2} \sinh ( \psi )\\
\frac {2} {3} \sinh ( \psi ) & \cosh ( \psi )
\end{array}
\right)
: \psi \in \mathbb {R}
}\punctsp.
\end{equation}
The boundary condition induced by $S$ is
\begin{equation}
g_0 = \sigma
\left(
\begin{array} {cc}
9 & 0\\
0 & 4
\end{array}
\right)\punctsp,
\end{equation}
where $\sigma$ will become the unknown constant conformal factor.

We are set now to apply Ricardo's formula. The intermediate calculations are straightforward yet cumbersome. The resulting metric does not always give back the connection of \eqrf{ExampleConnection}, but that was expected since not every value of $\xi_{ij}$ yields a metric connection. It can be proven that the following constrains make the connection a metric one:
\begin{align}
\xi_{12} &= \xi_{10} \, \xi_{20} \punctsp, \\
\xi_{21} &= \xi_{10} \, \xi_{20} \punctsp, \\
\xi_{11} &= - \alpha \, \xi_{20} + \xi_{10}^2 + \frac {9 ( \xi_{20}^2 - \xi_{22} ) } {4} \punctsp.
\end{align}

The components of the metrics compatible with the geodesics of \eqrf{ExParam} are
\begin{widetext}
\begin{multline}
g_{tt} = \sigma \Big[ - 9 - 36 \xi_{10} t - 18 \xi_{20} z + \frac {1} {2} \left( 8 \alpha^2 + 54 \alpha \xi_{20} - 180 \xi_{10}^2 - 81 \xi_{20}^2 + 81 \xi_{22} \right) t^2 - 4 \xi_{10} \left( 2 \alpha + 27
\xi_{20} \right) t z \\
+ \left( 4 \xi_{10}^2 - 18 \xi_{20}^2 - 9 \xi
_{22} \right) z^2 \Big] + \mathcal {O} \left( (t,z)^2 \right) \punctsp,
\end{multline}
\begin{multline}
g_{tz} = \sigma \Big[ - \left( 4 \alpha + 9 \xi_{20} \right) t + 4 \xi_{10} z + \left( - 16 \alpha \xi_{20} + 12 \xi_{10}^2 - 18 \xi_{20}^2 - 9 \xi_{22} \right) t z - 2 \left( 5 \alpha \xi_{10} + \beta + 18 \xi_{10} \xi_{20} \right) t^2 \\
+ 16 \xi_{10} \xi_{20} z^2 \Big] + \mathcal {O} \left( (t,z)^2 \right) \punctsp, \text {and}
\end{multline}
\begin{multline}
g_{zz} = \sigma \Big[4 + 8 \xi_{10} t + 16 \xi_{20} z + \left( - 8 \alpha \xi_{20} + 12 \xi_{10}^2 - 9 \xi_{22} \right) t^2 + 48 \xi_{10} \xi_{20} t z + 8 \left( 4 \xi_{20}^2 + \xi_{22} \right) z^2 \Big] + \mathcal {O} \left( (t,z)^2 \right)\punctsp.\\
\end{multline}
\end{widetext}


\end{document}